\documentclass[fleqn,10pt]{wlscirep}
\usepackage[utf8]{inputenc}
\usepackage[T1]{fontenc}

\title{1.5 million materials narratives generated by chatbots}

\author[1,2,$\dag$]{Yang Jeong Park}
\author[3,$\dag$]{Sung Eun Jerng}
\author[1,]{Jin-Sung Park}
\author[1,]{Choah Kwon}
\author[4,]{Chia-Wei Hsu}
\author[4,]{Zhichu Ren}
\author[2,*]{Sungroh Yoon}
\author[1,4,*]{Ju Li}
\affil[1]{Massachusetts Institute of Technology, Department of Nuclear Science and Engineering, Cambridge, 02139, USA}
\affil[2]{Seoul National University, Department of Electrical and Computer Engineering, Seoul, 08826, Republic of Korea}
\affil[3]{Suwon University, Division of Construction Environment Energy, Suwon, 18323, Republic of Korea}
\affil[4]{Massachusetts Institute of Technology, Department of Materials Science and Engineering, Cambridge, 02139, USA}

\affil[*]{corresponding author(s): Sungroh Yoon (sryoon@snu.ac.kr), Ju Li (liju@mit.edu)}

\affil[$\dag$]{these authors contributed equally to this work}

\begin{abstract}

The advent of artificial intelligence (AI) has enabled a comprehensive exploration of  materials for various applications. However, AI models often prioritize frequently encountered materials in the scientific literature, limiting the selection of suitable candidates based on inherent physical and chemical properties. To address this imbalance, we have generated a dataset of 1,494,017 natural language-material paragraphs based on combined OQMD, Materials Project, JARVIS, COD and AFLOW2 databases, which are dominated by ab initio calculations and tend to be much more evenly distributed on the periodic table. The generated text narratives were then polled and scored by both human experts and GPT-4, based on three rubrics: technical accuracy, language and structure, and relevance and depth of content, showing similar scores but with human-scored depth of content being the most lagging.  The merger of multi-modality data sources and large language model (LLM) holds immense potential for AI frameworks to help the exploration and discovery of solid-state materials for specific applications.

\end{abstract}
\begin{document}

\flushbottom
\maketitle

\thispagestyle{empty}

\section*{Background \& Summary}\label{BS}  

Materials are of such significance in human history that the designations assigned to each era of civilization are predicated upon the prevalent materials of the time. With the advent of the climate crisis, the 21st century has presented humanity with a multitude of challenges, prompting the exploration of novel materials for diverse new applications (solar cells\cite{fthenakis2009sustainability, ibn2017perovskite}, batteries\cite{cheng2021perspective, larcher2015towards, tan2020nanoscale},  catalysts\cite{ding2019carbon, baena2019carbon}, etc.) in {\em as short time as possible} in order to wean the {\em entire economy} off burning fossil fuels.   The expeditious discovery of materials possessing desirable attributes for specific applications garners considerable attention; however, it is impeded by the lack of digestible information (to a mechanical or electrical engineer,for example), both quantitative and qualitative, about materials. For example,  when asked about a specific material "${\rm Li_4Mn_5Ni(PO_4)_6}$", even a materials expert would usually turn to Google search, and the outcome would be quite varied and  dense literature with no guarantee of finding what one wants, that can take hours or days to parse through, which is just too slow, especially if all one needs is an initial screening.  Oftentimes, it is hard to present aggregated information, as properties are spread over multiple experimental and ab initio databases.

The desired properties (figure-of-merit) required to realize a given specific device may be known while the specific materials embodying superior figure-of-merit are generally unknown and difficult to identify. Throughout history, materials with technological functionalities have frequently been discovered through a combination of intuition, trial and error, and fortuitous circumstances. Today, the prevailing paradigm has transitioned towards a more comprehensive exploration of the vast space of potential materials. This endeavor is facilitated by the application of first-principles calculations and AI techniques. Notably, the advent of generative AI models has spurred a surge of research  into the realm of inverse material design\cite{noh2020machine, wei2019machine, so2020deep}. Through the utilization of generative AI techniques, researchers have been able to accelerate the process of materials discovery and design, offering promising opportunities for breakthroughs in the figure-of-merit for specific applications. Some of the authors have also examined the utilization of automated systems capable of generating scientific hypotheses in their recent work\cite{park2023can}. These systems based on large language model (LLM), including chatbots such as ChatGPT\cite{ChatGPT}, possess an inherent probabilistic nature that enables them to generate intriguing hypotheses, thereby expediting scientific advancements akin to human researchers. However, the  examples presented in Supplementary information section \ref{SIsec1} also demonstrate certain challenges with the "common-core" LLMs such as the standard ChatGPT\cite{park2023can}, including bias toward "hot materials" and "hot topics", whereas true ground-breaking innovations may spring from "cold topics" or less well-known materials. The "common-core" LLMs, owing to its learning process based on the probabilistic distribution of tokens, tends to prioritize the presentation of materials frequently encountered on the web and in scientific literature and publications
\cite{liu2018challenges, samson2019bird, zheng2018review, hou2020challenges, chen2018sulfide}, rather than "comprehending" in the human expert way the inherent properties and structures of materials and selecting suitable candidates more rationally. This is because the "common-core" text corpus found on the web are highly tilted toward materials already studied by human researchers, which can be rather limited, as researchers tend to flock toward "hot materials". This may limit the inventiveness of the narratives and inferences generated directly with "common-core" ChatGPT\cite{park2023can}.  This work aims to generate more balanced plain-language materials narratives that can be supplemented to the common corpus and used to {\em further} train more specialized LLMs so their inferences will be less biased toward "hot" but narrow-based materials.

In recent years substantial progress has been made in the realm of multimodal learning across diverse domains. The amalgamation and integration of information from various modalities, encompassing text, images, audio, and video, have facilitated breakthroughs in comprehending intricate data. This interdisciplinary approach has yielded remarkable applications in computer vision, natural language processing (NLP), and audio analysis, thus empowering the development of more comprehensive and resilient learning systems. But the field of materials research has yet to embrace the endeavor of multimodal learning. To surmount these challenges, our research team has  generated and shared data of 1,494,017 natural language-material pairs utilizing publicly available material databases and chatbots. This is a fairly large number considering that the number of data in ImageNet is 1,281,167.

The fusion and convergence of multiple modalities to enhance learning and comprehension of materials represent a relatively uncharted territory. However, given the rapid advancements in machine learning and the increasing availability of multimodal datasets, this captivating area of study harbors considerable potential for future research and innovation.
Our textual narratives will serve as an initial stepping stone towards pioneering novel subfields of artificial intelligence, such as materials captioning, materials multimodal learning, and simulation automation.

\section*{Methods}
\subsection*{Materials imbalances in common corpus}
\begin{figure}[ht]
\centering
\includegraphics[width=\linewidth]{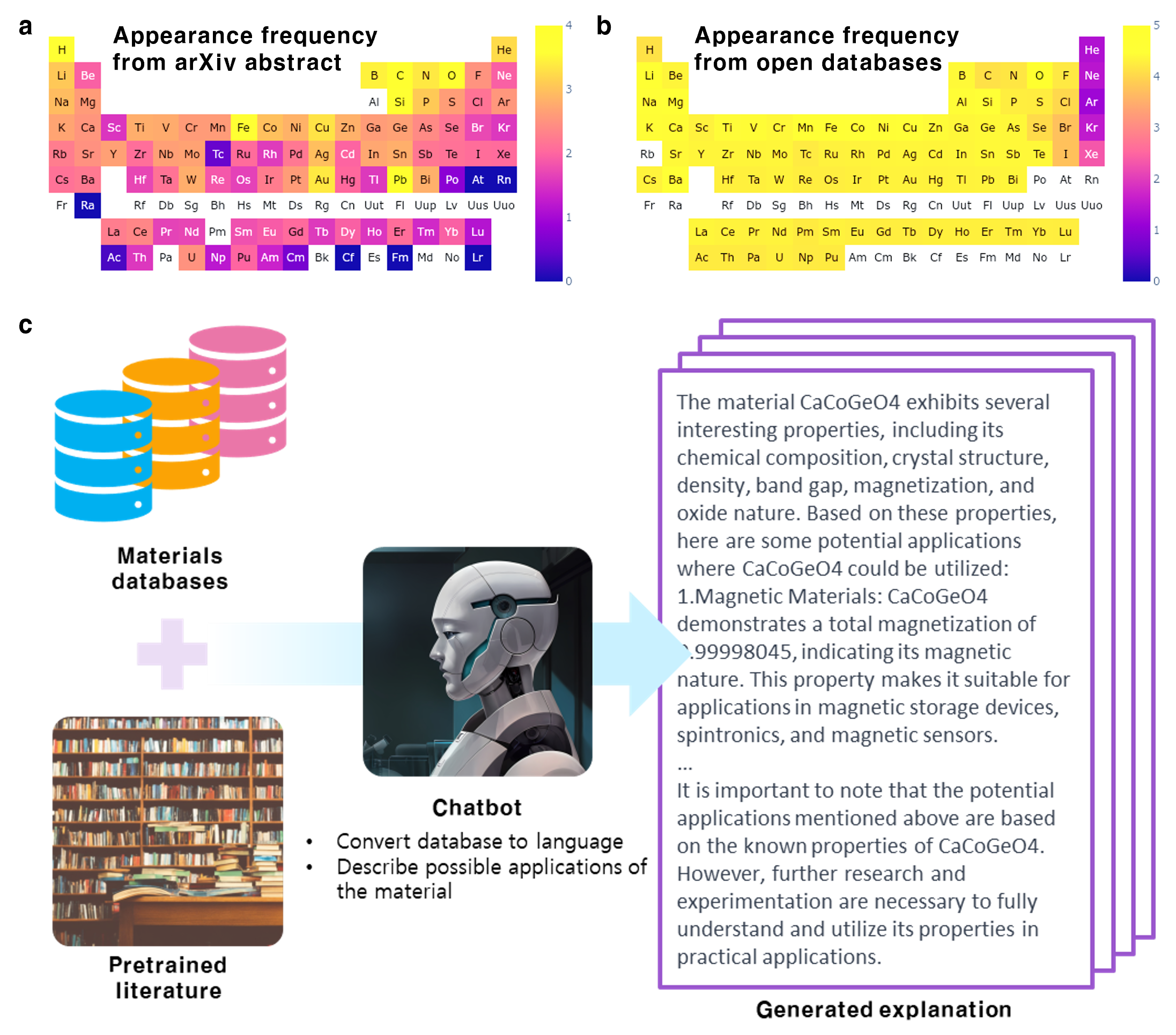}
\caption{Frequency distribution of chemical elements invoked in materials studied in a) the materials research literature, and b) publicly accessible databases that are mostly generated by {\em ab initio} calculations. The graph illustrates the bias or imbalance in materials of interest in a), focusing on oxides and frequently encountered materials like iron and copper. In contrast, the distribution of materials in b), excluding noble gases, is more evenly distributed, in open databases such as Materials Projects, JARVIS, and Open Quantum Material Database (OQMD). c) The discrepancy between the materials studied in research and those available in public databases, demonstrates the potential of language models to extract knowledge about materials and connect them with databases.}
\label{fig:Figure1}
\end{figure}

We visualize the bias present in the distribution of  materials described in the common-core text corpus, which for ChatGPT\cite{ChatGPT} are "array of sources available on the internet prior to September 2021. This includes a diverse range of documents, websites, books, and other text-based sources". To identify patterns of material bias found in actual academic literature, we utilized the joint automated repository for various integrated simulations (JARVIS)\cite{choudhary2020joint}-ArXiv dataset. Specifically, we selected abstracts from 284,815 papers in the `cond-mat' category. In order to identify the frequency of appearance of a material, as the chemical space is rapidly enlarged when a material of more than binary system is included, the frequency of occurrence was extracted by searching for a matching pattern using a regular expression for each element. We then extracted and visualized the occurrence frequencies by searching for matching patterns using regular expressions for each element. At the same time, the appearance frequencies of elements included in materials stored in publicly accessible databases such as Materials Projects\cite{MaterialsProject}, JARVIS\cite{choudhary2020joint}, and Open Quantum Material Database(OQMD)\cite{OQMD} were extracted and visualized. As shown in Figure \ref{fig:Figure1}, the materials studied within the research community focus on oxides, with a high occurrence frequency of familiar materials such as iron and copper. In contrast, most chemical elements (excluding noble gases) are much more evenly distributed in materials addressed by {\em ab initio} databases. This means that if one could combine the knowledge presented in specialized {\em ab initio} databases with a "common-core" LLM\cite{park2023can}, one could produce more balanced narratives that can be used to {\em further} train more specialized LLMs so their inferences will be less biased toward "hot" but narrow-based materials.  With such 
more specialized LLMs, one could extrapolate trained information of language models from the scientific literature. For instance, a language model can extract the fact that a material possessing an appropriate bandgap, high electrical conductivity, and stability can be considered a potential semiconductor candidate.

\subsection*{Material narrative text generation}
\begin{figure}[ht]
\centering
\includegraphics[width=\linewidth]{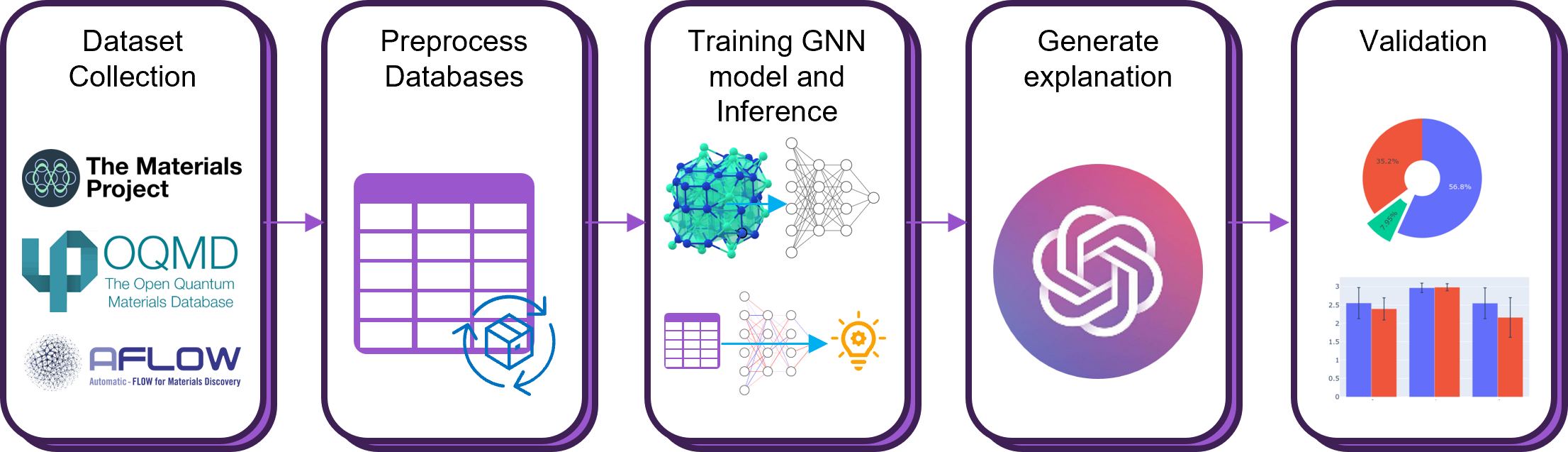}
\caption{Summary of the materials narrative generation process. The pipeline involved data collection from the joint automated repository for various integrated simulations (JARVIS). The databases were preprocessed to select relevant properties for textual description generation. A graph neural network (GNN) model was trained to extrapolate properties across multiple databases, and the results were saved as a Pandas DataFrame. The generated description went through two stages: first, converting the DataFrame into a dictionary and requesting a description, and second, using the generated result to obtain the final description. The generated descriptions were evaluated by human experts and GPT-4, and a validation process was conducted to evaluate correctness and detect potential adverse effects.}
\label{fig:Figure2}
\end{figure}

The process of generating the description of materials is summarized in Figure \ref{fig:Figure2}. Data collection pipeline was mainly implemented using Python programming language (version 3.9.15) and PyTorch\cite{paszke2019pytorch} (version 2.0.0), widely used in deep learning. All the computations were performed on a high-performance workstation with specifications including Intel® Core™ i9-10920X X-series Processor and NVIDIA RTX3090 graphic processing units (GPUs).\\

1. Dataset Collection: We obtained material data from publicly available repositories, the joint automated repository for various integrated simulations (JARVIS)\cite{choudhary2020joint}. The dataset encompassed diverse materials and covered a wide range of density functional theory (DFT) calculated properties. Moreover, JARVIS also provides an integrated way to access other publicly available databases such as Materials Projects and OQMD. The selected databases and number of materials included are described in Table \ref{tab:nummat}.\\

2. Preprocess Databases: To generate a textual description, some of the properties provided by each published database were heuristically selected. For example, scalar physical quantities such as "band gap" and "formation energy per atom", categorical data such as "crystal system", and Boolean data such as "stable" were mainly selected. The number of materials for which the properties were provided for each open database is summarized and shown in the table. The types of attributes provided are inconsistent and the number of types is different. For example, some properties such as "band gap" and "formation energy per atom" are provided by several databases, but some properties such as scintillation attenuation length are only provided by AFLOW2. \\

3. Training GNN model and Inference: Databases such as COD provide a large number of materials but no properties, making it difficult to generate meaningful descriptions on their own. Therefore, it was extrapolated using a graph deep learning model to create descriptions with an equal amount of information regardless of the source database. The model was modified to be E(3) equivariant based on ALIGNN\cite{choudhary2021atomistic}, which was successful in predicting quantum chemical properties. \\

The selected GNN model was implemented using deep learning frameworks, PyTorch and Deep Graph Library (DGL)\cite{wang2019deep}. The AdamW optimizer with normalized weight decay of $10^{-5}$ was used. A learning rate reduction strategy during plateaus was employed and training was conducted for 500 epochs with early stopping applied if no improvement was observed. All the training process was carried out using an NVIDIA RTX 3090 24 GB GPU. The model was trained on high-performance computing systems equipped with powerful GPUs.  See the table for the performance indicators of each model. The trained model was used to extrapolate each property with materials from multiple databases as input. The extrapolated results were saved as a Pandas DataFrame. The training results for each model are in Table \ref{tab:example}. For the regression task, a MAD:MAE ratio of 5 or greater is considered to make significant predictions, and some features fall slightly below that with values of 4 or greater, but we did not discard them. Traits with a MAD:MAE ratio of 4 or less have been removed.\\

4. Generate narrative: Creating the narrative was done in two stages. First, the data frame obtained by extrapolation was converted into a dictionary and requested as follows.\\

    `\textit{The following dictionary contains the composition and properties of a material stored in the database. Please write a description of the material, referring to this information. Make sure not to omit any item, and include all numerical values, citing their units appropriately. Feel free to include brief explanations or qualitative meanings for each property}.' + \textit{dictionary of material}\\

After that, the generated result was used as input again to obtain a final description.\\

    `\textit{Let's assume that we have a material with the following properties. Provide possible application areas for this material and explain the rationale behind them.}' + \textit{generated text}\\

This format shows a similar tendency to report new materials in actual academic papers. It is meaningful to follow a similar format as most researchers report the properties of a new material first and then list possible applications from it.\\

5. Validation: Evaluate whether the resulting material description is correctly described and free of other potential adverse effects. All descriptions were generated with ChatGPT-3.5-turbo and evaluated by human experts and GPT-4. In addition, it was investigated whether it was possible to identify whether the generated contents were written by artificial intelligence. A more detailed process is described in Validation Section.

\section*{Data Records}
The 2.3 million pieces of natural language-material paragraphs generated via a chatbot in this work will be deposited on Figshare after peer review. The database is organized in Pandas DataFrame format where elements in each column represent the same contents such as chemical properties, chemical formula, or generated text, and elements in the same row relate to the same material. The column data descriptions are listed in Supplementary information Table S1.

\section*{Validation}
Like any narrative from any source, ours will also contain factual errors and soft inaccuracies. The key is to reduce these as much as possible.

\begin{figure}[ht]
\centering
\includegraphics[width=\linewidth]{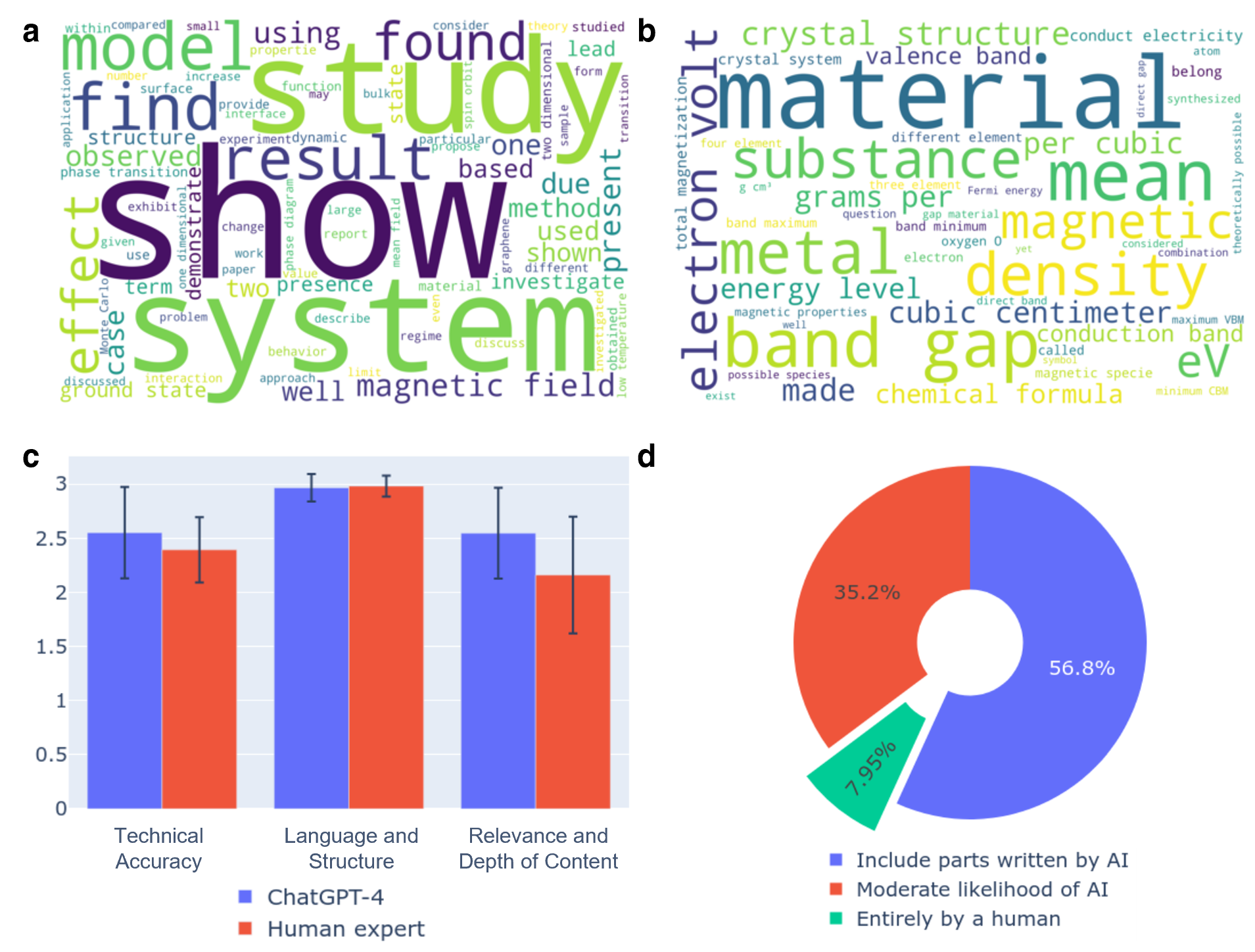}
\caption{The evaluation of the quality of the text generated by the Word Cloud visualization. Subfigure a) shows the highlighted words in the JARVIS-arXiv dataset, indicating the materials studied and their specific results. Subfigure b) visualizes descriptions generated from databases, often including possible applications based on stored material properties. c) To assess the generated text, evaluation criteria based on technical accuracy, language and structure, and relevance and depth of content were used. A random sample of 1,067 texts was evaluated by both human experts and GPT-4, showing similar results. d) To address concerns of content contamination, GPTZero was utilized to detect the AI-written text, achieving over 92\% accuracy in classifying the generated text.}
\label{fig:Figure3}
\end{figure}

The quality of the text generated by the Word Cloud visualization was evaluated in Figure \ref{fig:Figure3} a and b. A Word Cloud is a visual indicator of the frequency and importance of text, helping us to identify key themes and emphasized words in whole text. Through this, it was possible to evaluate how diverse and meaningful the generated texts were and how faithful they were to the main theme. In the JARVIS-arXiv dataset, all the input text is an abstract, so the corresponding word is highlighted to indicate that the material was studied and produced a specific result. In common, since each material in the generated description shows an almost uniform element distribution, it has a relatively low frequency of appearance, so it is not visualized in the word cloud. On the other hand, descriptions generated from databases are often visualized with descriptions of possible applications based on the stored material properties.

It is impractical to manually validate the hundreds of thousands of generated sentences individually. Since ChatGPT-3.5-turbo has already demonstrated its ability to generate natural-sounding sentences, traditional metrics used in NLP such as BLEU, ROUGE, and perplexity scores, which quantify similarity, coherence, and fluency of generated sentences, are not suitable for evaluating scientific and academic writing. Therefore, we have automated the evaluation using GPT-4 based on the following rubrics. (Figure \ref{fig:Figure3} c)\\

1. Technical Accuracy: The first and most crucial step is to evaluate the factual accuracy of the article. As the article is related to material science, it should properly represent scientific theories, facts, experimental observations, and material properties. This involves cross-referencing the information in the article with reliable sources to ensure accuracy.\\

2. Language and Structure: This evaluates how the AI has organized and presented the information. Is the article logically structured? Are the sentences well-formed and free of grammatical errors? Does the language use meet the standard of a scientific paper or article? The language should be clear and precise, and the information should be organized in a coherent and easy-to-follow manner.\\

3. Relevance and Depth of Content: This step examines whether the AI-generated content stays on topic and goes into enough depth. It should not merely scratch the surface of the subject but delve into the complexities and nuances. Also, the AI should not drift away from the topic or include irrelevant information.\\

To ensure statistical representativeness, we randomly selected 1,067 paragraphs from the 2.3 million  narratives generated using a random seed of 42 for evaluation. This selection accounts for a margin of error of plus or minus 3 percent at a 95 percent level of confidence. 
The selected texts were evaluated and compared by human experts as well as GPT-4. The evaluation results showed similar results in human experts and GPT-4. The texts are well organized, based on the database, and grammatically and structurally almost perfect. However, it is noteworthy that the human expert group gave a rather low score for the depth of the content.

Contamination of content created using generative AI by mixing it with the original content is one of the  challenges facing the large language model (LLM) community\cite{mitchell2023detectgpt, tian2023gptzero}. It is important to recognize the risks that the textual narratives generated by our method will be distributed indiscriminately as "100\% factual" and get mixed with human-generated text, polluting the corpus and hindering the progress of science and technology. In this context, various sensing technologies have emerged to prevent contamination and prevent indiscriminate usage. We used GPTZero\cite{tian2023gptzero} to assess the risk of our generated text going undetected. The results in Figure \ref{fig:Figure3} were correctly classified as over 92\% AI-written text, mitigating the risk somewhat, but still such risk is present.

\section*{Usage Notes}
The natural language text-material paragraphs created here can serve as a new starting point for LLM-based inverse material design to discover functional materials in silico, linking the efforts of the NLP  and  materials science communities. Examples of possible approaches for inverse engineering techniques using this database are as follows: 

1. Text-graph multimodal learning and inference of materials. By using NLP, it is possible to identify a subset of initial material structures with desired characteristics and desired application fields and convert it into actual first-principles calculation input through tools such as pymatgen\cite{pymatgen} and ASE\cite{ase}.

2. Fine-tuning large language models for scientific purposes. It is expected that large-scale applications of scientific hypothesis machines\cite{park2023can} can be achieved by fine-tuning large language models for specific purposes, based on domain-specific databases.

3. Vector database and use for in-context learning. Due to the emergent abilities\cite{wei2022emergentabilities} of large pre-trained language models, it is expected that vocabulary used in material fields can be understood "naturally" if a sufficiently large model is used.

Furthermore, we believe that the intelligence-driven "fifth paradigm"\cite{zubarev2019cognitive, leng2023fifth} of material discovery can be further advanced by introducing an approach that goes beyond mere data provision. Instead of relying solely on the knowledge embedded within the chatbot, we propose a methodology that generates text within a given context through connectivity with external databases. This approach has the potential to contribute significantly to the maturation of the fifth paradigm by expanding its scope beyond the mere provision of data.

\section*{Code availability}
Code for downloading generated descriptions will be available after peer-review.

\bibliography{manuscript}

\section*{Acknowledgements} 

This work was supported by a grant from the National Research Foundation of Korea (NRF) funded by the Korean government, Ministry of Science and ICT (MSIT) (No. 2021R1A6A3A01086766). The 05-Neuron supercomputer was provided by the Korea Institute of Science and Technology Information (KISTI) National Supercomputing Center. JL acknowledges support by DTRA (Award No. HDTRA1-20-2-0002) Interaction of Ionizing Radiation with Matter (IIRM) University Research Alliance (URA).

\section*{Author contributions statement}

Y.J.P., J.L. conceived the experiment(s), Y.J.P, S.E.J. conducted the experiment(s), Y.J.P., J.S.P, C.K., C-.W.H., Z.R., and S.Y. analysed the results. All authors reviewed the manuscript.

\section*{Competing interests} (mandatory statement)

The corresponding author is responsible for providing a \href{https://www.nature.com/sdata/policies/editorial-and-publishing-policies#competing}{competing interests statement} on behalf of all authors of the paper. This statement must be included in the submitted article file.

\newpage

\newpage

\newpage

\newpage
\begin{table}[ht]
\centering
\begin{tabular}{|l|r|}
\hline
Database name & Number of materials \\
\hline
JARVIS & 55,723\\
\hline
Materials Project & 126,335\\
\hline
OQMD & 460,046\\
\hline
Aflow2 & 420,135\\
\hline
COD & 43,1778\\
\hline
Total & 1,494,017\\
\hline
\end{tabular}
\caption{\label{tab:nummat}Number of materials used in this work.}
\end{table}

\newpage
\begin{table}[ht]
\centering
\begin{tabular}{|l|l|l|l|l|l|}
\hline
Property & Units & Trained from & MAD & MAE & MAD:MAE \\
\hline
Total energy per atom & eV/atom & OQMD (312,675) & 1.642 & 0.06307 & 26.03\\
\hline
Formation energy per atom & eV/atom & OQMD (312,670) & 0.6634 & 0.04511 & 14.71\\
\hline
Volume & Å³ & Materials Project (126,335) & 357.3 & 8.843 & 40.41 \\
\hline
Energy above hull & eV/atom & Materials Project (126,335) & 0.2055 & 0.0501 & 4.104\\
\hline
Band gap & eV & Materials Project (126,335) & 1.233 & 0.2484 & 4.963\\
\hline
Density & g/cm³ & Materials Project (126,335) & 2.182 & 0.0372 & 58.65\\
\hline
Enthalpy per atom & eV/atom & AFlow2 (420,135) & 1.732 & 0.0307 & 56.36\\
\hline
Scintillation attenuation length & cm & AFlow2 (420,135) & 0.8242 & 0.0186 & 44.36\\
\hline
\end{tabular}
\caption{\label{tab:example}Performance indicators of each model. The trained model was employed to extrapolate properties using materials from multiple databases as input. The extrapolated results were saved as a Pandas DataFrame. In the regression task, a MAD:MAE ratio of 5 or higher is deemed to indicate significant predictions. Although some features slightly fall below this threshold with values of 4 or higher, they were not discarded. Traits with a MAD:MAE ratio of 4 or less were removed.}
\end{table}

\newpage
\section*{Supporting Information}
\renewcommand{\thefigure}{S\arabic{figure}}
\setcounter{figure}{0}
\renewcommand{\thetable}{S\arabic{table}}
\setcounter{table}{0}
\setcounter{section}{0}
\section{Limitations of Causal Language Models in Material Selection}\label{SIsec1}
In this section, we asked ChatGPT-3.5 to create solid electrolyte materials for potential all-solid-state batteries:\\

`You are a great materials scientist. Please recommend potential solid electrolyte materials for the development of new all-solid-state batteries.'\\

The question was repeated three times, and a new agent was called every time instead of reusing the answer to prevent context contamination. Although the answers generated were different due to the stochastic nature of the generative model, it was found that certain materials appeared consistently. Among the commonly mentioned materials are: lithium garnet, such as Li7La3Zr2O12 (LLZO)\cite{liu2018challenges, samson2019bird}, perovskite-type oxides\cite{zheng2018review}, Sodium Super-Ionic Conductors (NASICON) materials, such as Na3Zr2Si2PO12 (NZSP) and NZPSO\cite{hou2020challenges}, or Sulfide-based materials like LGPS and LPS\cite{chen2018sulfide}.

It is important to note that the responses may not be exhaustive, and the field of solid-state batteries is evolving, with ongoing research introducing new materials and advancements. Staying updated with the latest scientific literature is crucial to keep track of the most recent developments in the field of all-solid-state batteries.\\

\color{blue}
Q1: You are a great materials scientist. Please recommend potential solid electrolyte materials for the development of new all-solid-state batteries.

A1: Thank you for the compliment! Solid-state batteries are an exciting area of research, and there are several potential solid electrolyte materials that show promise for their development. Here are a few recommendations:

1. Lithium Garnet (Li7La3Zr2O12 or LLZO): Lithium garnet is a popular solid electrolyte material due to its high ionic conductivity and stability with lithium metal electrodes. It has a wide electrochemical stability window and exhibits good mechanical properties. LLZO has been extensively studied and shows promise for high-performance solid-state batteries.

2. Perovskite-Type Oxides: Perovskite-type oxides, such as LaAlO3, LaGaO3, and Li0.33La0.557TiO3, have attracted attention as solid electrolyte materials. They exhibit high ionic conductivity and good stability with lithium. However, they may suffer from limited mechanical stability, requiring additional optimization and stabilization strategies.

3. NASICON-Type Materials: Sodium Super-Ionic Conductors (NASICON) materials, like Na3Zr2Si2PO12 (NZSP), have been investigated for solid-state batteries. They offer good ionic conductivity, stability, and are more abundant and less expensive compared to lithium-based materials. Sodium-ion solid-state batteries utilizing NASICON-type materials show promise for future energy storage applications.

4. Sulfide-Based Materials: Sulfide-based materials, such as Li10GeP2S12 (LGPS) and Li7P3S11, have shown high ionic conductivity and good stability with lithium electrodes. Sulfide-based solid electrolytes typically have higher ionic conductivity than oxide-based materials but may require careful handling due to their air and moisture sensitivity.

5. Composite and Hybrid Materials: Composite and hybrid solid electrolyte materials can combine different components to enhance overall performance. For example, a composite of LLZO and a polymer matrix can provide both high ionic conductivity and improved mechanical properties. Hybrid electrolytes, combining solid and liquid components, can also offer high conductivity while maintaining improved safety.

It's worth noting that the field of solid-state batteries is rapidly evolving, and ongoing research may introduce new materials or advancements. Therefore, it's essential to stay up-to-date with the latest scientific literature and technological developments in this area.\\

Q2: You are a great materials scientist. Please recommend potential solid electrolyte materials for the development of new all-solid-state batteries.

A2: Thank you for your kind words! All-solid-state batteries are a promising technology for next-generation energy storage. They offer several advantages over traditional liquid electrolyte batteries, such as improved safety, higher energy density, and wider operating temperature range. In the development of all-solid-state batteries, the choice of solid electrolyte materials is crucial. Here are a few potential solid electrolyte materials that have shown promise:

Lithium Garnet-Type Materials: Lithium garnet materials, such as Li7La3Zr2O12 (LLZO), Li6.25Al0.25La3Zr2O12 (LLAZ), and Li6.75La3Zr1.75Ta0.25O12 (LLZTO), are widely studied for all-solid-state batteries. They exhibit excellent chemical and electrochemical stability, high ionic conductivity, and good compatibility with lithium metal anodes.

Perovskite-Type Materials: Perovskite oxides, including various compositions of ABO3, have attracted attention as solid electrolytes. Examples include La0.5Li0.5TiO3 (LLTO), Li3xLa2/3-xTiO3 (LLTO), and BaZrO3-based materials. Perovskite electrolytes often exhibit high ionic conductivity at elevated temperatures and can be doped or modified to enhance their performance.

Sulfide-Based Materials: Sulfide-based electrolytes, such as Li10GeP2S12 (LGPS), Li7P3S11 (LPS), and Li3PS4 (LPSI), have demonstrated high ionic conductivity at room temperature. They offer good compatibility with lithium metal and can operate at lower temperatures compared to oxide-based electrolytes. However, sulfide electrolytes may have challenges related to stability and compatibility with other cell components.

NASICON-Type Materials: Sodium superionic conductor (NASICON) materials, such as Na3Zr2Si2PO12 (NZSP) and Na3Zr2PSi2O12 (NZPSO), have shown promising ionic conductivity and good stability. While initially investigated for sodium-ion batteries, they can also be explored for all-solid-state lithium batteries.

Polymer Electrolytes: Solid polymer electrolytes offer flexibility, ease of processing, and potential for lower manufacturing costs. Polyethylene oxide (PEO) and its derivatives, combined with lithium salts, are commonly used polymer electrolyte systems. Strategies to enhance their conductivity and stability, such as ceramic fillers or polymer composites, are actively pursued.

It's important to note that the development of all-solid-state batteries is an active research area, and new materials are continuously being explored. The performance and suitability of these materials depend on various factors, including ionic conductivity, stability, interface compatibility, and processing techniques. Further research and optimization are necessary to overcome the existing challenges and enable commercial viability of all-solid-state batteries.\\

Q3: You are a great materials scientist. Please recommend potential solid electrolyte materials for the development of new all-solid-state batteries.

A3: Thank you for your kind words! All-solid-state batteries have gained significant attention due to their potential to enhance battery safety, energy density, and cycle life. Several solid electrolyte materials have been investigated for their suitability in all-solid-state batteries. Here are some promising options:

Lithium Garnet (Li7La3Zr2O12 or LLZO): Lithium garnet is a widely studied solid electrolyte material. It exhibits excellent ionic conductivity and stability with lithium metal anodes. LLZO has a high Li-ion conductivity at room temperature and good compatibility with lithium electrodes.

NASICON (Na3Zr2Si2PO12 or NZSP): NASICON is a sodium superionic conductor that has shown promise as a solid electrolyte for sodium-based all-solid-state batteries. It offers good ionic conductivity and chemical stability, making it suitable for high-performance sodium-ion batteries.

Perovskite Oxides: Perovskite-type oxides, such as Lanthanum Gallate (La0.9Sr0.1Ga0.8Mg0.2O3 or LSGM) and Lanthanum Strontium Titanate (La0.5Sr0.5TiO3 or LST), have been investigated for their potential as solid electrolytes. They exhibit high ionic conductivity at elevated temperatures and can be compatible with lithium or sodium electrodes.

Sulfide-based Materials: Various sulfide-based compounds, such as Li10GeP2S12 (LGPS) and Li7P3S11 (LPS), have attracted attention as solid electrolytes for all-solid-state batteries. These materials offer high Li-ion conductivity and have the advantage of being compatible with lithium metal electrodes.

Glass/Ceramic Composites: Composite materials combining glass and ceramic components can offer a balance between ionic conductivity and mechanical stability. For example, sulfide glass-ceramics like Li2S-P2S5-LiI have shown promising results for solid electrolytes due to their good ionic conductivity and improved stability compared to pure sulfide materials.

It's important to note that the field of solid electrolyte materials is continuously evolving, and ongoing research may reveal new candidates with enhanced properties. Therefore, it's advisable to consult the latest scientific literature and ongoing research to stay up-to-date with the most recent developments in the field of all-solid-state batteries.\\

\color{black}
\section{Examples of generated descriptions}\label{SIsec2}
In this section, we will present 10 examples of generated material descriptions for understanding.\\

\textbf{1. Generated description about $SbSBr_3F_6$}\\
The material with formula $SbSBr_3F_6$ is stored in the database. The substance has a formation energy per atom of -1.50833 eV/atom, which suggests that the material can form through an exothermic process. It has a band gap of 2.35 eV, which is an indication of its potential use in electronic devices.   The total energy per atom of the substance is -1.79239 eV/atom and it has an energy above hull of 0.122655 eV/atom, which shows that the material is stable but not perfectly stable. It has a space group symbol of $P2_1/c$, indicating its symmetry and crystal structure, which is monoclinic.   The density of the material is 3.71 g/cm³ and its volume is 908.9747 Å³, which indicates that it is a relatively dense substance. The total magnetization per formula unit is 0.298496 $\mu B/f.u.$, which suggests that it may exhibit magnetic properties.   The enthalpy per atom of the material is -3.622568 eV/atom, indicating its potential use in energy storage. The scintillation attenuation length of the substance is 3.490982 cm, indicating its potential use in detecting radiation.   The material is classified as "not oxide" and is currently unstable. In summary, $SbSBr_3F_6$ is a dense, potentially magnetic, and semiconducting material that has potential applications in electronic devices, energy storage, and radiation detection. One potential application area for this material is in electronic devices that require a semiconducting material with a band gap of 2.35 eV. The high density of the substance can also make it suitable for use in electronic devices that require compact components. Another potential application area is in energy storage devices due to its high enthalpy per atom. This material can potentially be used as a cathode or anode in batteries, which could offer higher energy storage capacities compared to other materials.  With a scintillation attenuation length of 3.490982 cm, this material also has potential applications in detecting radiation. It can be used as a radiation detector in medical imaging or nuclear power plants.   Although it is not perfectly stable, the formation energy per atom of -1.50833 eV/atom indicates that the material can form through an exothermic process. Therefore, it may be possible to synthesize a more stable form of the material through further research and experimentation.  Lastly, the potential magnetic properties of the substance can also be investigated. If its magnetic properties are harnessed, it can be used in magnetic storage devices such as hard drives or memory cards.  In summary, $SbSBr_3F_6$ is a promising material with applicable properties in various fields such as electronics, energy storage, radiation detection, and potentially in magnetic storage devices. Its high density makes it ideal for compact components and its formation energy per atom suggests it can be synthesized through further research.\\

\textbf{2. Generated description about $K_3AuSe_13$}\\
The substance is $K_3AuSe_13$, which has a chemical formula consisting of 3 potassium atoms, 13 selenium atoms, and 1 gold atom. The formation energy per atom is -0.41144 eV/atom, indicating that this material can be formed spontaneously with a low amount of energy required. The band gap, which is the energy difference between the highest occupied energy level and the lowest unoccupied energy level of the material, is 0.652 eV. This indicates that $K_3AuSe_13$ is a semiconductor material, which can be used in devices such as solar cells. The total energy per atom is -1.64329 eV/atom, which represents the amount of energy required to remove an atom from the material. The energy above hull is 1.168405 eV/atom, which shows that it is not the most stable configuration of atoms. $K_3AuSe_13$ has a space group symbol of P2/c and a crystal system of monoclinic, indicating the geometric arrangement of the atoms. Its density is 4.501 g/cm³, and its volume is 999.16797 Å³. The total magnetization is 0.298495 $\mu B/f.u.$, which represents the total magnetic moment of the material per formula unit. The enthalpy per atom is -3.600214 eV/atom, which is a measure of the heat released or absorbed by the material during a process. The scintillation attenuation length is 2.74214 cm, which indicates the distance over which the light intensity is reduced by a factor of e due to energy absorption of photons by the material. $K_3AuSe_13$ is not an oxide-type material, and it is not stable in its current configuration. $K_3AuSe_13$ can be a suitable material for applications that require semiconductors. It has a band gap of 0.652 eV, which is considerably low, and hence it can absorb light efficiently. This property makes it feasible for use in solar cells or photovoltaic devices as an absorber layer to convert light into electrical energy. Furthermore, $K_3AuSe_13$ has a scintillation attenuation length of 2.74214 cm, which indicates that it can stop the attenuation of light, making it advantageous for use in scintillation detectors such as radiation detectors.   Although $K_3AuSe_13$ is not a stable material, its spontaneous formation with a low amount of energy required and its moderate density of 4.501 g/cm³ makes it a feasible option for use in chemical sensors. Lastly, $K_3AuSe_13$ has a total magnetic moment of 0.298495 $\mu B/f.u.$. and a monoclinic crystal system, which makes it worthwhile for applications that require magnetic materials, such as data storage, magnetic field sensors, and magnetic resonance imaging (MRI) diagnostics.\\

\textbf{3. Generated description about Mg}\\
The material with the chemical formula Mg is a stable substance with a cubic crystal system and Im-3m space group symbol. Its density is 1.792 g/cm³, and its volume is 23.080276 Å³. The formation energy per atom is 0.03228 eV/atom, which indicates that the material is stable. Its band gap is zero eV, meaning it is a metal and does not have a band gap. The total energy per atom is 1.16522 eV/atom, while the energy above hull is also 0.03228 eV/atom. The substance’s total magnetization per formula unit is 0.29849 $\mu B/f.u.$. and its enthalpy per atom is -1.506615 eV/atom. The scintillation attenuation length of the material is 6.544316 cm. The substance is not an oxide material, and it is characterized as a non-oxide type.Based on the properties of the material provided, here are some possible application areas and reasons for them:  1. Aerospace Industry: Magnesium is a lightweight metal, so it may be used as an attractive material in the aerospace industry for various applications such as construction of lightweight aircraft structures, engine components, and missile components.  2. Medical Industry: Magnesium is biocompatible and also has good biodegradability characteristics which make it suitable for various medical applications such as implants, stents and drug delivery systems.  3. Automotive Industry: The low density of magnesium makes it a promising candidate to be used as a replacement for heavy materials in the manufacturing of automotive parts such as engine blocks, transmission cases, and wheels, which improves fuel efficiency and reduces overall weight.  4. Energy Industry: Magnesium is a good thermal conductor, so it can be useful in the field of energy generation and storage. Its properties may be utilized in the development of batteries, fuel cells, and heat exchangers.  5. Defense Industry: Magnesium alloys have been used at high levels in the defense industry for various military applications because of its high strength-to-weight ratio and other physical properties. It is used for armor plating, missile parts, and aircraft components.  Overall, Magnesium metal has diverse potential uses owing to its excellent physical properties.\\

\textbf{4. Generated description about $Ni_2Mo$}\\
The substance described here has the formula $Ni_2Mo$, made up of nickel and molybdenum elements. The formation energy per atom is -0.10411 eV, which indicates that this material is relatively stable. The substance has no band gap, indicating that it is a metal. The total energy per atom is -3.68128 eV, while the energy above hull is 2.350084 eV/atom, indicating that the substance is not thermodynamically stable. The space group symbol is Immm, and the crystal system is orthorhombic. The density of the substance is 9.555 g/cm³, the volume is 36.561306 Å³, and the total magnetization is 0.298496 $\mu B/f.u.$ The enthalpy per atom is -7.543179 eV/atom, and the scintillation attenuation length is 1.256536 cm. The substance is not an oxide, and it is not stable.Based on the properties of this material, it may have potential applications in the following areas:  1. Catalysis: Nickel-molybdenum alloys are known for their exceptional catalytic properties and are commonly used in a wide range of industrial processes. The stability and metallicity of this material make it a potential candidate in industries such as petroleum refining, ammonia synthesis, and hydroprocessing.  2. Magnetic storage: The substance has a total magnetization of 0.298496 $\mu B/f.u.$, indicating that it could have potential applications in magnetic storage devices such as hard drives.  3. Radiation detection: The material's scintillation attenuation length is 1.256536 cm, indicating its potential in radiation detection technologies such as radiation dosimeters and imaging systems.  4. Alloy development: The properties of $Ni_2Mo$ make it a suitable candidate for alloying with other elements to improve the mechanical, thermal, and chemical properties of the resulting material. It can be used in the manufacture of high-strength alloys for applications such as aircraft engines, gas turbines, and nuclear power plants.  In conclusion, the properties of $Ni_2Mo$ make it a promising candidate for applications in catalysis, magnetic storage, radiation detection, and alloy development. Further research is required to explore these potential applications and to optimize the material's properties for specific purposes.\\

\textbf{5. Generated description about $KNiIO_6$}
The substance in question has the chemical formula $KNiIO_6$. It forms with a total energy per atom of -2.96134 eV/atom and an energy above hull of 1.77847 eV/atom. Its formation energy per atom is -0.85383 eV/atom.   The substance has a band gap of 1.196 eV, which is the amount of energy required to move an electron from the top of the valence band to the bottom of the conduction band. The scintillation attenuation length of the substance is 2.722038 cm, which means it can absorb and re-emit light.  The space group of the substance is symbolized by P312, with a crystal system of trigonal. It has a density of 4.09 g/cm³, with a volume of 129.62222 Å³. The substance has a total magnetization of 0.872899 $\mu B/f.u.$, indicating its magnetic properties. Its enthalpy per atom is -4.552144 eV/atom, which is the heat generated or absorbed in chemical reactions.  The substance is an oxide, and it is unstable (stable: False).The substance in question is a rare-earth metal oxide known as potassium nickelate. Given its unique properties, there are a number of possible application areas for this material, including:  1. Photovoltaics: With a band gap of 1.196 eV, potassium nickelate could be utilized for solar energy conversion. The material could potentially absorb light in the visible and near-infrared range, generating electron-hole pairs that could be collected to produce an electrical current. The scintillation attenuation length of 2.722038 cm should also make it efficient at converting light to electrical energy.  2. Magnetic storage: With a total magnetization of 0.872899 $\mu B/f.u.$, potassium nickelate could be utilized for magnetic storage. For example, the material could be used in the fabrication of hard disk drives, where data is stored magnetically. The magnetic properties of the material would allow it to be magnetized and demagnetized easily, enabling data to be written and read.  3. Catalysis: Given its enthalpy per atom of -4.552144 eV/atom, potassium nickelate could act as a catalyst to facilitate chemical reactions. The heat generated or absorbed in chemical reactions could be used to drive forward or reverse reactions, respectively. For example, the material could be used to facilitate the conversion of biomass into biofuels by breaking down complex organic molecules into simpler components.  4. Magneto-optical devices: With its unique optical and magnetic properties, potassium nickelate could be utilized in the fabrication of magneto-optical devices. For example, the material could be used in the production of magneto-optical sensors, which are used for magnetic field sensing. Magneto-optical devices could also be used to manipulate light using magnetic fields.  Overall, the properties of potassium nickelate make it a promising candidate for a variety of applications, ranging from solar energy conversion to magnetic storage and catalysis. Its unique combination of optical and magnetic properties makes it particularly appealing for use in magneto-optical devices.\\

\textbf{6. Generated description about $La_12C_6I_17$}\\
The substance stored in the database has the chemical formula $La_12C_6I_17$. It has a monoclinic crystal system with a C2/c space group symbol. Its volume is 2589.194378 cubic units, and its density is 4.997669 grams per cubic centimeter. The energy per atom of this material is -5.248304 electron volts, and its formation energy per atom is -1.233994 electron volts. The energy above hull is 0 eV, indicating that it is in a stable state. The substance is not an oxide. Its band gap is 0 eV, implying it is a metal or zero-gap semiconductor. The total magnetization of the material is 0.944722 Bohr magnetons. The enthalpy per atom is -5.04623 eV, and its scintillation attenuation length is 2.718653 cm. Overall, this material is a dense and stable metal or zero-gap semiconductor with a monoclinic crystal structure composed of lanthanum, carbon, and iodine.Possible application areas for this material are as follows:  1. Nuclear Reactors: Due to its high density, this material can be used as a radiation shielding material in nuclear reactors to protect workers from ionizing radiation. Additionally, its scintillation attenuation length suggests that it could be used in scintillators for radiation detection.  2. High-Temperature Applications: Lanthanum and carbon are both high-temperature resistant materials, suggesting that this material could be used in high-temperature applications such as furnace linings or as a heat shield in spacecraft.  3. Electronic Devices: The zero-gap semiconductor property of this material may make it useful in electronic devices, particularly in the development of efficient solar cells. Its high magnetization may also prove useful in the development of spintronic devices.  4. Catalysis: Lanthanum-based compounds have previously demonstrated potential as catalysts in chemical reactions, and the inclusion of carbon and iodine in this material may improve its catalytic properties.  Overall, the high density, stability, and unique properties of this material make it an interesting candidate for a range of applications in various fields, including nuclear engineering, materials science, electronics, and catalysis.\\

\textbf{7. Generated description about $VBiO_4$}\\
The material stored in the database has a chemical formula of $VBiO_4$, with an oxide type composition. It has an orthorhombic crystal system with a space group symbol of Pbcn. The material's density is 6.503184 g/cm3 and has a volume of 330.841633 angstroms cubed.   The energy per atom is -6.849485 eV, while the formation energy per atom is -2.137913 eV. The material is stable and has a total magnetization of 0.000012 Bohr magnetons at 0K. The energy above hull is 0.020202 eV, indicating that the material is nearly stable with respect to decomposition into other materials.   The band gap of the material is 2.5804 eV, which means that it is a semiconductor with a relatively low band gap. The enthalpy per atom is -6.456616 eV, which indicates that the material is exothermic and gives off energy when it forms.   Finally, the scintillation attenuation length of this material is 1.167155 cm, which is a measure of how far light can penetrate the material before being absorbed or scattered. Overall, this material has interesting properties that make it worthy of further study for potential applications in electronics, optics, and energy storage.The material, $VBiO_4$, with a relatively low band gap, could potentially be used in electronic devices as a semiconductor material. Its stable nature and exothermic formation energy also make it a viable option for energy storage applications.  The scintillation properties of this material, with a moderate attenuation length, suggest its use in radiation detection or imaging applications.  Additionally, its magnetic properties could make it useful in spintronic applications where magnetic materials are used to store and manipulate data.  The orthorhombic crystal system and space group symbol of Pbcn make this material an interesting candidate for research in crystallography and solid-state physics, which could lead to further understanding and development of novel materials with similar properties.  Overall, the combination of its various properties makes $VBiO_4$ an intriguing material for potential applications in electronics, optics, energy storage, radiation detection, imaging, and solid-state physics. Further research in these areas can reveal more potential applications for this material.\\

\textbf{8. Generated description about $SiO_2$}\\
The substance in question is a crystalline material with a monoclinic crystal system and a space group symbol of C2/m. Its chemical formula is $SiO_2$, suggesting that it is an oxide. The material has a density of 1.658787 g/cm³ and a volume of 721.773352 Å³. Its total magnetization is 0.0 and it is stable. The energy per atom is -7.907814 eV and the formation energy per atom is -3.269574 eV. The energy above hull is 0.008424 eV and the enthalpy per atom is -6.284725 eV. The material has a band gap of 5.6611 eV, indicating that it is an insulator. Finally, the scintillation attenuation length of the material is 6.081382 cm, meaning that it is able to absorb and re-emit light over a short distance. Based on the properties provided, the material described is most likely alpha-quartz, which has a wide range of potential applications due to its distinct physical and chemical properties. Here are some possible application areas for this material:  1. High-Temperature Applications: Alpha-quartz has a high melting point of 1713°C and exhibits excellent thermal stability. Therefore, it can be used as a refractory material for high-temperature applications, such as furnace linings in the steel industry.  2. Semiconductor Industry: The wide bandgap of alpha-quartz makes it an excellent insulator and a suitable material for electronic components in the semiconductor industry. Its unique properties also make it useful as a substrate material in the production of silicon wafers.  3. Optics Industry: Alpha-quartz has excellent transparency in the UV range, making it ideal for applications in the optics industry, including lenses, windows, and prisms.  4. Chemical Industry: Alpha-quartz is highly resistant to chemical corrosion and can be used as a container material for corrosive chemicals. It can also be used as a filter medium due to its high porosity.  5. Medical Imaging: Alpha-quartz has a high scintillation attenuation length, meaning that it can absorb radiation energy and efficiently emit light. This property makes it a suitable material for use in medical imaging applications, such as scintillation detectors in CT scanners and gamma cameras.  In summary, alpha-quartz has unique properties that make it useful in a variety of industries, including high-temperature applications, the semiconductor industry, optics, the chemical industry, and medical imaging. Its thermal stability, optical transparency, chemical resistance, and scintillation properties make it a versatile and valuable material for many applications.\\

\textbf{9. Generated description about $Ti_2VSe_4$}\\
The substance stored in the database is $Ti_2VSe_4$. It has a monoclinic crystal system and belongs to the C2/m space group symbol. The substance has a volume of 137.203097 cubic units and a density of 5.597724 grams per cubic centimeter. The energy per atom is -6.586902 and the enthalpy per atom is -6.61834, both expressed in electron volts. The formation energy per atom is -1.036029 eV, while the energy above the hull is 0.015983 eV. The scintillation attenuation length value is 2.30469 centimeters. $Ti_2VSe_4$ is not an oxide and is considered stable. Additionally, the substance has a total magnetization of 1.268687 Bohr magnetons and a zero band gap, which means it is a metallic material.$Ti_2VSe_4$ is a metallic material, which means it is an excellent conductor of electricity. It also has a total magnetization, making it a magnetic material. These properties make $Ti_2VSe_4$ suitable for various applications, such as:  1. Magnetic storage devices: The material's magnetic property makes it an excellent candidate for use in magnetic storage devices. $Ti_2VSe_4$ can be used in hard drives, magnetic tapes, and similar devices that require materials with good magnetic properties.  2. Catalysis: $Ti_2VSe_4$ has been studied for its catalytic properties. It has shown excellent performance in catalyzing various reactions, including the reduction of CO2. Its excellent conductivity and magnetic properties could also be advantageous in electrocatalytic processes.  3. Superconductors: Although $Ti_2VSe_4$ is not a superconductor, it can be used in the fabrication of superconductors. This is because its electronic structure and density of states are compatible with known superconducting materials.  4. Thermoelectrics: $Ti_2VSe_4$ has a high Seebeck coefficient, which is a measure of a material's ability to convert temperature differences into electrical voltage. This property makes it useful in thermoelectric applications, such as waste heat recovery.  In summary, $Ti_2VSe_4$ is a metallic material with excellent magnetic and electrical conductivity properties, making it well suited for a range of applications including magnetic storage devices, catalysis, superconductors, and thermoelectrics. Its stability and compatibility with other materials further enhance its potential applications.\\

\textbf{10. Generated description about $Li_4Mn_5Ni(PO_4)_6$}\\
The substance with the pretty formula $Li_4Mn_5Ni(PO_4)_6$ is a triclinic crystal system oxide with a space group symbol P1. Its density is 3.262262 g/cm³, and the volume is 473.880239 angstroms cubed. The formation energy and energy per atom are -2.467747 eV and -7.043534 eV, respectively, while the enthalpy per atom is -6.208705 eV. The energy above hull is 0.06303 eV, and the band gap is 0.4639 eV. The total magnetization is 24.994525 µB, and the scintillation attenuation length is 3.857658 cm. This substance is stable. $Li_4Mn_5Ni(PO_4)_6$ is a promising material with potential applications in energy storage and conversion, catalysts, and scintillators.   1. Energy storage and conversion: Lithium ion batteries are widely used in portable electronic devices, electric vehicles, and grid-scale energy storage due to their high energy density, long cycle life, and low self-discharge rate. $Li_4Mn_5Ni(PO_4)_6$ is a cathode material that can replace conventional cathodes, such as LiFePO4 or LiCoO2, due to its higher energy density and better rate capability. The material's high energy above hull, low formation energy, and band gap make it a promising candidate for energy storage applications. The material could also potentially be used in solid-state batteries since it is a stable oxide.  2. Catalysis: The surface area of $Li_4Mn_5Ni(PO_4)_6$ is relatively large, leading to an increased reaction rate, which can be useful in catalysis. Additionally, the material is stable, making it a good candidate for use in catalytic applications. The material's high enthalpy per atom and total magnetization may also play a role in catalytic reactions.  3. Scintillators: Scintillators are materials that emit light when exposed to ionizing radiation, making them useful for radiation detection. The large scintillation attenuation length of 3.857658 cm and high total magnetization make $Li_4Mn_5Ni(PO_4)_6$ a promising candidate for use in scintillators.  In summary, $Li_4Mn_5Ni(PO_4)_6$ has properties that make it a stable, versatile material with potential applications in energy storage and conversion, catalysis, and scintillators. Further research and development could lead to innovative applications in the future.\\

\newpage
\begin{table}[ht]
\centering
\begin{tabular}{|l|l|l|l|l|l|l|}
\hline
Property & OQMD & Materials Project & JARVIS & COD & AFLOW2 \\
\hline
Spacegroup symbol &  & 126,355 & 55,723 & & \\
\hline
Crystal system &  & 126,355 & 55,723 & & \\
\hline
Total energy per atom & 312,675  & 126,355 & 55,723 & & 420,135\\
\hline
Formation energy per atom & 312,670  & 126,355 & 55,723 & & \\
\hline
Volume &  & 126,355 & & & \\
\hline
Energy above hull &  & 126,355 & 55,723 & & \\
\hline
Band gap  &  & 126,355 & 55,723 & & \\
\hline
Density  &  & 126,355 & 55,723 & & \\
\hline
Total magnetization  &  & 126,355 & & & \\
\hline
Oxide type  &  & 126,355 & & & \\
\hline
Enthalpy per atom &  & & & & 420,135\\
\hline
Scintillation attenuation length  &  & & & & 420,135\\
\hline
\end{tabular}
\caption{\label{tab:S1} Data columns used in this work.}
\end{table}

\end{document}